\documentclass[12pt]{article}
\usepackage{graphicx}
\usepackage{latexsym} 

\begin{document}
 
\begin{center}
{\bf {\huge Interpreting Mathematics in Physics: Charting the Applications of SU(2) in 20th Century Physics}}\\
\hspace{10pt}\\
Ronald Anderson\footnote{Email: ronald.anderson@bc.edu} \\
{\em Department of Philosophy, Boston College,} \\
{\em Chestnut Hill, MA 02467, U.S.A.}\\
 \hspace{10pt}\\
G. C. Joshi\footnote{Email: joshi@physics.unimelb.edu.au}\\
{\em School of Physics}\\ 
{\em University of Melbourne,}\\
{\em Victoria 3010, Australia}\\
\hspace{10pt}\\
April 28, 2006
\end{center}
\hspace{10pt}\\
 
 \begin{abstract}

\indent The role mathematics plays within physics has been of sustained interest for physicists as well as for philosophers and historians of science.  We explore this topic by tracing the role the mathematical structure associated with SU(2) has played in three key episodes in 20th century physics -- intrinsic spin, isospin, and gauge theory and electro-weak unification.  We also briefly consider its role in loop quantum gravity.  Each episode has led to new physical notions of a space other than the traditional ones of space and spacetime, and each has had associated with it a complex and in places, contested history.  The episodes also reveal ways mathematical structures provide resources for new physical theorizing and we propose our study as a contribution to a need Roger Penrose has identified to develop a ``profoundly sensitive aesthetic'' sense for locating physically relevant mathematics.  
\end{abstract}

\section{Introduction}

Early in the 20th century Whitehead observed that as mathematics increasingly entered into ever greater extremes of abstract thought it became at the same time increasingly relevant for the analysis of particular concrete facts (\cite{whitehead1926}, p. 47).  With a similar focus on the significance of developments in mathematics for physics, Dirac proposed in various publications that a powerful method of advancing physics was one that drew on the resources of pure mathematics by attempting  ``to perfect and generalize the mathematical formalism that forms the existing basis of theoretical physics, and after each success in this direction, to try to interpret the new mathematical features in terms of physical entities \ldots '' \cite{dirac1931}.  In addition, for Dirac, the beauty of the mathematics also plays an important role in guiding search for successful physical theories.  Similar concerns to do with identifying features of the mathematics most likely to be of use in mathematical physics have been raised by Roger Penrose in his recent book {\em The Road to Reality, A Complete Guide to The Laws of The Universe} \cite{penrose2005}.  For Penrose, the quality at issue requires a certain aesthetic sensitivity:

\begin{quote}

\ldots progress towards a deeper physical understanding, if it is not able to be guided in details by experiment, must rely more and more heavily on an ability to appreciate the physical relevance and depth of the mathematics, and to `sniff out' the appropriate ideas by use of a profoundly sensitive aesthetic mathematical appreciation (\cite{penrose2005}, p. 1026).

\end{quote}

Penrose's book itself may be taken as an impressive contribution to such a project. The general manner in which the role of mathematics in physics has often been discussed, however, tends to obscure locating those  features of the mathematics most appropriate for advances in physics.  The issue though is a crucial one, not simply for understanding the nature of mathematical physics, but also for the actual practice of physics given the increasing complexity of the mathematical structures within physical theories.    

By tracing closely particular episodes in the history of physics we propose that new perspectives can emerge on this topic.  This was the approach of an earlier study by the authors, where the generalization implicit in the Division Algebras, in the sequence from real to complex numbers and to quaternions, was explored as an exemplification of Dirac's thesis that generalizations of mathematical structures currently successful in physical theorizing provide resources for new physical theories \cite{anderson1993}.   Here we seek to explore the relationship in another such focused manner by considering a number of episodes where a particular basic mathematic structure, the two dimensional special unitary group SU(2), has been used in physics.  Three in particular -- intrinsic electron spin, isotopic spin or isospin, and Yang-Mills gauge theory and electroweak unification --  involved new degrees of freedom and generalized notions of ``spaces'' and were accompanied by a complex history where new ideas were contested and only stabilized gradually.  That forms of space are at issue touches topics of contemporary relevance given the way in which string theories require more than three spatial dimensions.  The fourth episode, loop quantum gravity, forms a research program underway, one that also entails a profound transformation of our traditional notions of space as the continuous ``background'' in which events occur, one that rivals that of Einstein's general theory of relativity where spacetime was given a dynamical significance.  The spirit of the points we wish to raise is in accord with El Naschie's recent article (also exploring an observation of Penrose on the place of mathematics in physics) on the physical significance of the mathematics of transfinite set theory, computability and fractal geometry \cite{elnaschie2005}.

\section{Intrinsic Spin}

Our first example where the mathematics associated with SU(2) emerged as a crucial part of the description of a new physical phenomena is the quantum mechanical description of the electron spin that emerged in the 1920s.  Accounts of this development have been given by Pais \cite{pais1986}, \cite{pais1991}, Tomonaga \cite{tomonaga1997}, and van der Waerden \cite{waerden1960}, as well as in reflections by various of the original participants: Pauli \cite{pauli1946}, Goudschmidt \cite{goudschmidt1971} \cite{goudschmidt1976}, and Uhlenbeck \cite{uhlenbeck1976}.  The experimental phenomenon of concern at the basis of this development was the anomalous Zeeman effect where a splitting of the spectral lines took place in the presence of a magnetic field (for the original work of Stern and Gerlach in 1922 on the splitting of spectral lines see Ref. \cite{gerlach1922}).

As Pauli later remarked in 1946, the unique type of splitting involved appeared in the early 1920s as ``hardly understandable'' and ``unapproachable'' (quoted in van der Waerden  \cite{waerden1960}, p. 200).  As a way to explain the effect Pauli invoked in 1924 a new quantum property for the electron that he referred to as ``two valuedness not describable classically.''  Then in the following year he proposed there existed a ``fourth degree of freedom of the electron''  and an exclusion principle that there can never be more than two electrons in an atom characterized by the same set of quantum numbers.  Later in his Nobel award address Pauli noted that one of the initial difficulties for physicists understanding the exclusion principle initially lay in the absence of a meaning ascribed to this fourth degree of freedom \cite{pauli1946}.  

R. de L. Kronig in January of 1925, aware of Pauli's extra degree of freedom \cite{pauil1925}, proposed the idea of an electron spinning about its axis as a solution to characterizing the extra degrees of freedom of the electron.  The notion though met with resistance at the time by Pauli, Heisenberg, and others with the result that Kronig did not publish it.  Then independently of Kronig, in October of 1925, although hesitantly, Uhlenbeck and Goudschmit proposed in a short paper the idea of a spinning electron \cite{goudschmidt1925}. They presented a fuller note in the journal {\em Nature} in February of 1926 with an added comment by Bohr containing his endorsement of the idea \cite{goudschmidt1926b} (see also Ref. \cite{goudschmidt1926a}).  A letter later in the year by L. H. Thomas, also in {\em Nature}, showed how an unresolved issue of a doublet splitting in the spectrum could be accounted for by a more careful treatment of relativistic effects of electron motion.  This converted Pauli to the notion of a spinning electron \cite{thomas1926}.  Pauli was to remark later that subsequent work by Bohr showing that the electron spin was indeed a non classical and essentially a quantum mechanical property of the electron confirmed his earlier sense of the necessity of a ``non describable classically'' phenomena \cite{pauli1946}.  

What is evident from the writings of the time, set against the backdrop of the establishment of the new quantum mechanics, and from later reminiscences, is the struggle to make sense of the new property of matter that spin represented.  Miller in particular, has traced the complex transformation of visualizability during this period, noting the essential non-visualizablity of Pauli's ``fourth degree of freedom'' (\cite{miller1984}, p. 139).  Schr\"{o}dinger's often quoted remark in 1926 about being repelled by Heisenberg's approach with its ``transcendental algebra'' and lack of visualizability is a pointer to these discussions.  On this topic Pauli subsequently provided a striking account of how understanding electron spin entailed accepting an abstract mathematical representation instead of visualizability in quantum mechanics.

\begin{quote}
After a brief period of spiritual and human confusion, caused by a provisional restriction to `Anschaulichkeit' [visualizability], a general agreement was reached following the substitution of abstract mathematical symbols as for instance psi, for concrete pictures.  Especially the concrete picture of rotation has been replaced by mathematical characteristics of the representations of the group of rotations in three-dimensional space (\cite{pauli1955}, p. 30).

\end{quote}

The relevant mathematical characterization was given by Pauli in 1927 \cite{pauli1927} and shortly afterwards in the same year by Darwin.  Pauli's paper, entitled in English ``Contribution to the Quantum Mechanics of the Magnetic Electron,'' introduced his famous three spin operators using Schr\"{o}dinger's new 1926 ``method of eigenfunctions.''  The operators Pauli presented as obeying the relations:

\begin{equation}
\displaystyle
s_{x}s_{y} = -s_{y}s_{x} = is_{z} \ldots, \:
s_{x}^{2} = s_{y}^{2} = s_{z}^{2} = 1
\end{equation}

In a footnote Pauli mentioned how Jordan had alerted him to the significance of these relations and their connection to quaternions, and thereby forging their association with a mathematical structure with a complex and rich history since its discovery by Hamilton in 1843.  Pauli also noted the quantization of the spin variable in a specified direction and the contrast with the classical situation. 

\begin{quote}

However, in contrast to classical mechanics, the variable can, quite independently from any special nature of the external force field, only take on values $1/2$ $h/2\pi$ and $-1/2 h$$/2\pi$. The appearance of this new variable produces, therefore, in an electron simply a splitting of the eigenvalues into two local position $\Psi_{a}$ and $\Psi-{b}$ and more generally in the case of N electrons of 2N functions; these are to be viewed as ``probability amplitudes'' that in a given stationary state of the system not only the position coordinates of the electrons fall within given infinitesimal intervals, but that the components of their eigenmomenta in the defined direction have for  $\Psi_{a}$  the given value $1/2$ $h/2\pi$ and for  $\Psi_{b}$  value $-1/2$$h/2\pi$. 

\end{quote}

Pauli also presented a matrix representation of the now familiar matrices associated with his name:
       
\begin{equation}
s_{x} = \pmatrix{0 & 1\cr 1 & 0} \: \: s_{y} = \pmatrix{0 & -i\cr i & 0} \: \: s_{x} = \pmatrix{1 & 0\cr 0 & -1}
\end{equation}

Pauli realized that this new form of angular momentum was quite different from classical angular momentum defined by position and momentum and such that it could not be turned off as in the classical situation.  The matrix representation given by Pauli was particularly important to him as with this representation he could relate his work to the quantum mechanics of Jordan and Heisenberg.

In the same year Darwin published a short article in {\em Nature} \cite{darwin1927a} on the ``spinning electron'' to be followed later in the year with a more detailed mathematical analysis \cite{darwin1927b}.  Darwin was writing in the light of having seen the proofs to Pauli's paper and his comments on Pauli form an interesting insight into early discussions on the best formalism with which to express spin.  While using quaternions to represent the spin states, Darwin claimed his mathematical analysis was similar to Pauli's, remarking though that Pauli was ``disposed to regard the wave theory as a mathematical convenience and less as a physical reality.''  For Darwin, an essential point of Pauli's analysis was representing the electron with two separate functions.  For Darwin the two components can be further interpreted more physically to form a vector wave function (later known as spinors) similar to the analysis of light.  Yet on this point he noted that his differences with Pauli were ``lying rather deep in the whole quantum theory'' (\cite{darwin1927b}, p. 244). Another feature of Darwin's paper is a strong statement that with the wave theory of electron spin visualization is now ``entirely lost'' (\cite{darwin1927b}, p. 230).

The significant feature that emerges from this complex history is that Pauli was able to present a mathematical formalism that could capture an entirely new phenomenon and concept in physics, one that provided a fusion of a relativistic explanation from Thomas and quantum mechanical notions.  Moreover, the years between the first emergence of the idea of electron spin and the mathematical analysis of Pauli and Darwin, from 1925 to 1927, were extraordinary years in physics with the birth of the new quantum theory (for an insightful comment on the controversies of this time see Ref. \cite{Kleppner2000}).  At that time the meaning of the notion of ``electron spin'' and the interpretation of the mathematical formalism used to capture the properties of spin, together with the issue of its visualizability, were all contested issues.  In his biography of Pauli, Enz notes the disagreement among historians concerning the introduction of spin, and the strangeness that ``in the history of modern physics the idea of spin has stirred up so much controversy'' ({\cite{enz2002}, p. 119). 

The group structure, SU(2), later associated with the Pauli matrices was to make clear the remarkable property of electron spin that quantum mechanically it takes two rotations of 360 degrees to return to the original state.  Mathematically this is related to SU(2) being a double cover of SO(3), the group of proper rotations in three dimensional space such that two elements of SU(2) correspond to one of SO(3).  Furthermore, while the Pauli matrices of SU(2) do not form a division algebra as do quaternions, there is an isomorphism between SU(2) and unit quaternions thereby tying together in an intriguing manner rotations in three dimensional space, quaternions and SU(2).  The representations of SU(2) also form spinors, originally discovered in a mathematical context by Cartan in 1913; objects with a fundamental role, perhaps more central that SU(2), in 20th century physics.  The classic text on gravitation by Misner, Thorne, and Wheeler nicely brings this cluster of ideas together \cite{misner1973}. Moreover, spin, as can be seen from the structure of the Pauli matrices, is associated with the central role imaginary numbers play in quantum theory \cite{baylis1992}, \cite{hestenes2003}.

In the year following papers by Pauli and Darwin, Dirac developed a quantum formalism for the electron consistent with relativity such to explain naturally the ``duplixity'' of spin phenomena \cite{dirac1928}.  The set of 4x4 matrices needed form a natural extension of the Pauli matrices and form a Clifford algebra.  They too are spinors and here spinors are linked with the Lorentz group.  Dirac's work stands as a striking example of how the mathematical richness of a physical equation (in this case the extra solutions of a manifestly covariant wave equation for the electron) provided resources for physical theorizing (for the prediction of an antiparticle).

\section{Isospin}

Heisenberg \cite{heisenberg1932} first introduced the Pauli matrices in 1932 in a three part article on the structure of the nucleus, which he took to be composed of protons and neutrons.  The neutron had just been discovered and although Heisenberg notes ``the neutron will be taken as an independent fundamental particle,'' he also considered the neutron as a possible composite of a proton and an electron, although an electron that needed to have zero spin that obeys Bose statistics (see Ref. \cite{bromberg1971} for an account of the history of the introduction of the neutron).  Heisenberg used the Pauli matrices to write a Hamiltonian function for the nucleus that included interactions between the neutron and protons and their kinetic energies.  Each particle in the nucleus was tagged by a ``fifth number'' that represented whether the particle is a proton or neutron, specified as the eigenvalue of the 3rd Pauli matrix.  The name Heisenberg used was ``$\rho$-spin'', where  $\rho$ represented the Pauli matrices. The move enabled the mathematical structure of spin states to be transposed to nucleon states.  At the time Heisenberg noted that ``of course'' the space spanned by them is  ``not the ordinary space'' yet when asked in an interview in 1979 about the meaning of ``internal symmetries'' such as isospin that do not appear as operations in the world, Heisenberg remarked rather elliptically, ``I suspect that isospin is a symmetry similar to space and time.  I cannot say that it is related to them'' \cite{buckley1996}.

Studies of the history of the concept of isospin such as those by L. M. Brown \cite{brown1988} and H. Kemmer  \cite{kemmer1982} have stressed that Heisenberg's use of the Pauli matrices was essentially a mathematical technique for keeping track of two particle states.  It is only with later developments that the group and symmetry properties associated with the charge independence of the nucleon interaction come to the foreground.  Indeed, Kemmer remarks that the Pauli matrix notation used by Heisenberg is largely a bookkeeping one in terms of  nucleon states and for this reason the use of ``such an elaborate and somewhat unusual formalism''  probably had ``little appeal'' to contemporary readers (\cite{kemmer1982}, p. 365).  Thus like the complexity in the emergence of the concept of electron spin, a similar complexity attends the origins of isospin.  For Pais, Heisenberg's papers represent a breakthrough simply in their insistence that the proton-neutron system within the nucleus be amenable to non-relativistic treatment of forces (\cite{pais1986}, p. 416).  As Brown's account makes clear, it was a series of papers in 1936 arguing for a charge independence of the nucleon interaction that solidified  the notion of the spin description as Pauli had invoked for electrons.  In particular, a paper by Cassen and Condon starts explicitly with an analysis in terms of spin matrices \cite{cassen1936}, and Wigner in 1937 elaborated such an analysis labeling such a degree of freedom as ``isotopic spin'' \cite{wigner1937}.  

Wigner introduced an operator to express the ``isotopic number'' for the nucleus, noting that while `` \ldots the mathematical apparatus of the isotopic spin is, hence, somewhat redundant'' at the same time it ``will turn out that it is very useful in spite of this'' (\cite{wigner1937}, p. 107).  In the same year, Kemmer applied the isospin formalism to Fermi's theory of the electron-neutrino field \cite{kemmer1937} noting that while the formalism could account for the symmetries of the problem, the magnitude of the forces could not be explained this way.  

Historically isospin is significant in being the first ascription of quantum numbers to elementary particles of a sort with no association with spacetime properties.  The concept of isospin was also significant in leading to the later SU(3) classification of hadrons in the three quark model that formed the ``Eightfold Way'' of Gell-Mann and NeÕeman.  Such ``flavor'' based classifications based on quark states dominated physics in the 1960s and 70s, contributing to a culture of group theory in particle physics of that time (for an insight into this group-dominated period of particle physics see a reminiscence by Lipkin in Ref. \cite{lipkin2002}).  The culture was such as to inspire higher order group theoretical explorations such as four types of quarks, and multiquark states of more than the standard three quarks states for baryons and two quark meson states (for an example of both topics reflecting the culture of the time, see a study by the authors in Ref. \cite{anderson1979}).  Moreover, the isospin symmetry is only an approximate symmetry, the flavor symmetry even more so, a notion that anticipates that of ``broken symmetry'' that dominates modern gauge field theory.    That the mathematics of SU(2) was one of the paths that opened out to the rich mathematical machinery of group theory through its generalization was of enormous significance for 20th century physics and illustrates another feature of the place of a mathematical structure in physics. 

\section{The Yang-Mills Gauge Invariance Paper and Beyond}

The central aspect of Yang and Mills's famous paper of 1954 relevant to our concern is their use of the notion of isotopic spin to develop a generalization of the gauge invariance of electromagnetism \cite{yang1954}.  Their article begins by noting that the notion of isotopic spin had recently been much discussed and gives a brief review of its history, from Heisenberg through to Wigner.  Yang and Mills present the conservation of isotopic spin as an invariance of all interactions under rotation in an isotopic spin space.  However, choosing whether a nucleon is a proton or neutron at a point is an arbitrary process and to allow that freedom independently at all spacetime points required their (now familiar) move of introducing a new field analogous to the electromagnetic field with a property of gauge invariance.  Yang and Mills go on to construct a set of field equations that satisfy gauge invariance.    
 
Implicit in this significant paper is a set of notions central in late 20th century physics: that  of a group space in which a gauge field take values, the idea of a gauge field as mediating interactions, and the notion of local invariance under ``rotations'' within the space of the gauge field.   As in the earlier two episodes where the mathematics of SU(2) had been invoked, problems arose about the physical relevance and significance of the resultant fields.  The first was immediately apparent -- that the new fields introduced are massless, a feature which Yang and Mills noted interfered with their interpretation as mediating nucleon interactions.  On this Yang was later to observe: ``If you read our article, you will see that we did not really know how to connect it with reality. And the main problem was with the mass of the gauge field. We knew that there was no charged massless gauge field. And yet we did not know how to put mass into it'' (\cite{yang1994}, p. 1446).  Also, Yang, in his commentary on the paper notes that in a seminar the previous year Pauli had pressed him on the issue of the mass of the gauge field (\cite{yang1983}, p. 20). 

 It was later with the concept of spontaneous symmetry breaking, developed in the early 1960s, where the mathematical symmetry remains but without a physical symmetry, and then in 1964, the realization by Higgs and others of a mechanism whereby massless Goldstone bosons that result from spontaneous symmetry breaking acquire mass, that the issue of mass was resolved.  The proof of the renormalizability of non-abelian gauge fields by 't~Hooft and Veltman in 1971 and 1972 completed the pieces that enabled local gauge theories to be a resource for accounting for interactions. 

Prior to the proof of renormalization, however, SU(2) emerged again, this time as part of the SU(2) x U(1) unified theory of electroweak interactions developed by Weinberg in 1967 and independently by Salam in 1968.  As a gauge theory, SU(2) generated three massless gauge bosons, and U(1) one such boson.  Through the Higgs mechanism the model generated three massive bosons that mediated the weak force, with a massless boson to correspond with the photon of electromagnetism.  The standard model extended this to include the SU(3) gauge group of massless gluons mediating the interaction between quarks.  The symmetry group is now transformed,  from SU(2) as representing quark states to a gauge symmetry.  The story of the rich developments of this period can be traced partly through the Nobel Prize address of Salam \cite{salam1980} and Weinberg \cite{weinberg1980}, and for a general history see Ref. \cite{oraifeartaigh2000}.

In this new context SU(2), as a symmetry of a gauge field, has indirectly become part of  a long discussion in philosophy of science contexts on the nature of gauge fields and how the mathematical structure of the gauge field maps to physical reality.  The issue here continues the extended discussion on the reality of potentials in classical electromagnetism, given a new dimension by Aharonov and Bohm in 1959 pointing out the physical significance of loop integrals of the gauge field in quantum contexts \cite{aharonov1959}. The property of gauge invariance of potentials has often been taken as indicating they are only part of the mathematical structure of the theory and not terms that represent physical phenomena directly.  The centrality of gauge fields within current physical theories, however, makes such an argument hard to defend.  An introduction to these issues may be found in Refs. \cite{belot2003} and \cite{healey2001}.

\section{Loop Quantum Gravity}

In our last example we briefly mention the approach for uniting general relativity and quantum theory known as loop quantum gravity.  The founders of this approach include Abhay Ashtekar, Lee Smolin, and Carlo Rovelli.   The approach assumes Einstein's theory can be quantized non-perturbatively.  The radical aspect of the approach lies in the transformation of conventional notions of space and time as a continuous background in which events occur, as classically understood, to a notion of spacetime as formed by loop-like states, which are essentially holonomies or structures formed by parallel transport along closed paths.  Holonomies have long featured in gauge field theories as gauge invariant quantities, representing curvature in the gauge space.  Within the approach of loop quantum gravity such holonomies become quantum operators. The role SU(2) plays in loop quantity gravity is as the gauge group of the field of the holonomies.  The choice by Rovelli and Smolin in 1995 of ``spin networks'' that Penrose had developed in the 1970s as a model of discrete quantum geometry (see: \cite{rovelli1995}, \cite{rovelli2004} and \cite{penrose2005}) was key in establishing these ideas.  While a spin basis network can be given for all compact gauge groups, the one that is relevant to quantum gravity is SU(2), in particular its spinor structure.  

The discrete nature of space arises in this approach from spin networks forming space directly.  Representations of SU(2) label the edges of the spin network in three dimensions, and in this way a mathematical description of the kinematics of a quantum gravitational field can be obtained in three spatial dimensions.  On its significance Rovelli remarks of the nature of a spin network: ``a spin network state is not in space: it is space.  It is not localized with respect to something else: something else (matterÉ) might be localized with respect to it'' \cite{rovelli2004}.  The evolution of a spin network in time, needed for a Feynman path integral formulation of the theory, generates a ``spinfoam'' structure.    For a clear introductory description of the group representations usually invoked here see Ref. \cite{nicolai2006}.

One of the consequences of this program (which still awaits completion) and the manner in which it has forged a place in competition to string theories, has been to raise deep conceptual questions about the nature of space and time (on this point see the review by Rickles \cite{rickles2005}).

\section{Conclusions}

The episodes sketched briefly above in the history of physics serve to illustrate the remarkable role the mathematical structure associated with SU(2) has played in modern physics, in particular to represent quantum phenomena.  In the context of the project of the paper to identify qualities of mathematical structures productive for the development of physics, we will draw three lessons.  

One central quality of such productive structures is illustrated by the way SU(2) provided the resources to represent new forms of spaces to capture properties of particles and their interactions, and in the case of loop quantum gravity,  to be woven into the basic features out of which space and spacetime is constructed.   Minimally SU(2) provided the mathematical structure to account for states with two degrees of freedom and to fully represent transformation between such states.  More sophisticated structures that have served such a role include Calabi-Yau manifolds, fiber bundles  and the general forms of phase and configuration space representations of dynamical systems.  The notion of representing  states of systems in this way is pervasive in contemporary physics.  The quality is a very general one, perhaps too general, but those mathematical structures that afford such possibilities play a crucial role in physics.   The group structure of SU(2) was key here, and in particular its spinor structure.  Group structures can express properties invariant in various spaces, another important quality for capturing physical properties.  The lesson here expresses one given by Dirac in 1939:  

\begin{quote}
It would probably be a good thing also to give a preference to those branches of mathematics that have an interesting group of transformations underlying them, since transformations play an important role in modern physical theory, both relativity and quantum theory seeming to show that transformations are of more fundamental importance than equations  (\cite{dirac1939}, p. 125).
\end{quote}

Moreover SU(2) is part of the remarkable story of the vital place of symmetries in 20th century physics.  The fundamental role of group structures in elementary particle physics has been one of the reasons some philosophers of physics have proposed that the ultimate reality consists of structural relations rather than the  entities these relations are between (see, e.g., French and Ladyman \cite{french2003}).

Another feature the above history reveals, one associated with the group structure in the case of SU(2), is the value for physics of a mathematical structure that can be generalized in various ways. The extended group products in the case of nucleon isospin states is an example of this feature as well as the ready generalization of SU(2) to higher unitary groups in terms of gauge field theories.  The quality of a productive mathematical structure is one of providing resources for the extension of physical theories and the generation of new ones.    Such qualities are ``theoretical'' virtues and here the use of higher dimensional spaces in string theories provide another example.  Whatever emerges on their physical significance they provide structures that allow possible theories to be formed embodying virtues such as unification. 

A third feature, although one we propose tentatively and one somewhat contrary to the previous virtue, is that some mathematical structures play key roles in physical theories by their constraints matching the constraints of nature.  The success of SU(2) for capturing spin phenomena and the role in loop quantum gravity is related to the three dimensional nature of space.  The quality here was perceptively identified by the 19th century physicist, P. G. Tait, when commenting on quaternions.  Tait (a fan of quaternions) noted that while to pure mathematicians quaternions have ``one grand and fatal defect'' that they cannot be applied to spaces of n-dimensions, from a physical point of view this defect is their ``greatest possible recommendation'' as they have the structure for the three dimensions in which we are ``doomed to dwell'' \cite{tait1871}.

The ``sensitive aesthetic mathematical appreciation'' that to Penrose identifies physically relevant mathematics will by nature entail a set of skills and qualities for the mathematical physicist that defy full articulation, yet attention to actual physical theories and their emergence and evolution, as our intent has been here, provides a way to pursue the issue.  The history we have sought to chart here suggests that attention to structures that capture the physical states of a system in a mathematically concise manner, that possess features that allow extensions through mathematical generalizations, and that possess constraints that match those of nature, are all qualities for the mathematical physicist to keep particularly in mind. 

We conclude with a proposal that the reverse direction to the one Penrose has presented is worth considering.  Namely, to suggest that exploring the manner in which certain physical and empirical conditions in general  are conducive for advances in mathematics, is a project that awaits  the mathematician and philosopher of mathematics.  On this point physics has long been a resource for the development of mathematics.  Moreover, the highly abstract and mathematical nature of contemporary physical theories suggests that there is reason to consider a blurring of the realms of mathematics and physics.  And affirming such a blurring Gregory Chaitin has recently maintained that the implications of G\"{o}del's incompleteness theorems for mathematics plus the use of computers to explore mathematics truths implies that mathematics may have a ``quasi-empirical'' quality \cite{chaitin2006}.  How then to develop the sort of sensitive aesthetic appreciation of those features of physics, and the particular contexts (such as the use of computers) in which contemporary mathematics is practiced,  that can identify those conducive for advances in mathematics?

 \subsection{Acknowledgments}  We are grateful to Zwi Barnea, Joy Christian, Maya Joshi, and Andy Martin for assistance on matters related to the paper.

 \end{document}